\documentclass[conference]{IEEEtran}
\IEEEoverridecommandlockouts

\usepackage{cite}
\usepackage{amsmath,amssymb,amsfonts}
\usepackage{algorithmic}
\usepackage{graphicx}
\usepackage{textcomp}
\usepackage{xcolor,inputenc}
\def\BibTeX{{\rm B\kern-.05em{\sc i\kern-.025em b}\kern-.08em
    T\kern-.1667em\lower.7ex\hbox{E}\kern-.125emX}}
\begin{document}

\title{Positioning and transmission in cell-free networks: ambiguity function, and MRC/MRT array gains\\
{\footnotesize 
}
\thanks{\footnotesize{Thomas Pairon is with AerospaceLab, Louvain-la-Neuve.} }
}

\author{\IEEEauthorblockN{
Luc Vandendorpe, Laurence Defraigne, Guillaume Thiran, Thomas Pairon and Christophe Craeye}
\IEEEauthorblockA{\textit{ICTEAM, UCLouvain, Louvain-la-Neuve, emails~: firstname.lastname@uclouvain.be} }
}
\maketitle

\begin{abstract}
Cell-free network is a new paradigm, originating from distributed MIMO, that has been investigated for a few recent years as an alternative to the celebrated cellular structure.
Future networks not only consider classical data transmission but also positioning, along the lines of Integrated Communications and Sensing (ISAC). The goal of this paper is to investigate at the same time the ambiguity function which is an important metric for positioning and the understanding of its associated resolution and ambiguities, and the array gain when maximum ratio transmission (MRT) or MR combining (MRC) is implemented for data communications. In particular, the role and impact of using a waveform with non-zero bandwidth is investigated. The theoretical findings are illustrated by means of computational results.  
\end{abstract}


\section{Introduction}
Cell-free networks are or have been the topic of many recent works \cite{emil-cf}. Focusing on the so-called user-centric cell-free network (UCCFN), the concept is to surround users by multiple access points (APs) (whose set is named cluster), either coherent or not, instead of the classical cellular structure where the APs are surrounded by multiple users. Each AP is itself equipped with either one or several antennas. 
These APs can collectively implement MRC from users in the uplink (UL), or MRT to users  in a downlink (DL) \cite{tse}. Furthermore it has been shown that thanks to the circular wavefront associated with such a distributed network, a more specific form of beamforming can be achieved which is named beampointing \cite{emil-mamimo}. It therefore enables Location based Division Multiple Access (LDMA) by means of which users are separated on the basis of their position. This is more appealing that the seminal Space Division Multiple Access (SDMA) technique where users can be separated thanks to their angle only, and which, by the way, would have been more appropriately named ADMA for Angle based Division Multiple Access. 

\medskip

A second function or service contemplated in future 6G wireless networks on top of data transmission is positioning \cite{henk1}. For positioning, a quantity enabling to understand the capabilities offered by the network, is the ambiguity function \cite{Woodward}.  It can be understood as the anticipation that for instance a maximum likelihood (ML) position estimator, equivalent to a least square (LS) estimator in AWGN conditions, will contain a correlation operation. This correlation operation can be tuned to a certain position but nevertheless have at its input the signal coming from another position. The ambiguity function is the output, in noiseless conditions,  of a correlator matched to a target user interfered by the signal originating from a second user. Such a function typically has an oscillatory shape for RF transmissions; it has a main lobe whose width provides information about the resolution, while the secondary lobes show how much the correlator will be sensitive to ambiguities  in noisy conditions. Positioning in cell-free networks has been investigated for APs equipped with radio-stripes \cite{stripes} operating in the far-field (FF). Performance bounds for positioning in UCCFNs have  been considered in \cite{henk2} as a component of future multi-service 6G networks equipped with ISAC capabilities. Paper \cite{polin1} has shown the improvement of communication, positioning and sensing metrics thanks to UCCFNs. 

\medskip

For downlink data transmission to a single-antenna user, a metric of interest is the so-called array gain. 
Similarly to the ambiguity function for positioning, the impact on  user 2 of the beampointer targeting user 1  is of interest to understand the resolution associated with the beampointer or the amount of interference  
likely to be generated on neighboring users. We here also investigate the array gain for MRT. In a dual setup, i. e. an uplink scenario where the APs operate MRC and when the receive beampointer is designed for a certain user, the array gain tells how much another user will generate interference on the user of interest \cite{emil-mamimo}. The analysis therefore also provides information about the separability of users.
A subsequent conclusion that  can be drawn from this analysis is the resolution with which the impulse response generated by an obstacle can be estimated. This aspect is left for future work.
To the best of the authors knowledge, analytical derivations for the ambiguity function and for the array gain achieved in UCCFNs offering positioning and communications services have not yet been reported.  Therefore, the following  contributions are identified for this paper:
\begin{itemize}
\item The ambiguity function achieved by a UCCFN operating positioning  is analytically investigated. Conclusions can be drawn about the resolution modeled by this ambiguity function, as well as the related ambiguities;
\item Data transmission is considered with, on the one hand, MRC in the uplink and, on the other hand, MRT in the downlink;
\item It is shown how the conclusions reported for the positioning ambiguity function apply to the array gains achieved in downlink or uplink;
\item Those investigations are extended to non-zero bandwidth transmitted signals as it is of course required for data transmission. The positive impact of such a signal on the ambiguities is proved.
\item These theoretical findings are illustrated by means of computational results.
\end{itemize}

\section{System model}
 
We  consider the situation where an antenna array is made of a circle of single antenna APs around targets here named user equipments (UEs). The circle has a radius $R$. With respect to the center of the circle, the user of interest has Cartesian coordinates $(x_s,y_s)$ or polar coordinates $(R_s,\theta_s)$. We will use the simplified expression of ``$(R_s,\theta_s)$-user` for the user having a position described by $(R_s,\theta_s)$. He/she is assumed to be positioned close to the center of the circle, and that we therefore have  $R  \gg R_s$.

\medskip
In the next sections we will consider the services of positioning and communications, and derive the ambiguity functions as well as the array gains. 
\section{Positioning and Communications}
 
\subsection{Positioning}

\subsubsection{Received signal}

We  first consider an infinite number of single antenna APs, meaning a space-continuous array. The distance between an AP located in $(x,y)$, or $(R,\theta)$, and the $(R_s,\theta_s)$ UE of interest is
 $R_{ds}$ and can in this circular scenario be approximated as
 \begin{eqnarray}
 R_{ds}
 &=&
 [(x-x_s)^2+(y-y_s)^2]^{0.5} 
 \nonumber 
 \\
 &=&
 R[1+R_s^2/R^2 - 2 R_s\cos(\theta-\theta_s)/R]^{0.5}
\nonumber 
 \nonumber \\
&\simeq&
 R - R_s\cos(\theta-\theta_s)
\label{cf1}
 \end{eqnarray}
where  we have used the order-1 Taylor series expansion of $\sqrt(.)$. 
We also define $z=R \theta$. 
with $\theta\in[0,2\pi]$. We assume that the APs and users are perfectly time and carrier synchronous.
The signal received at position $z$ can be written (we ignore the attenuation which can here be assumed to be identical for all antennas)
\begin{eqnarray}
 r(z,t;s)
 &=&
 \exp(-jk(R-R_s\cos(\theta-\theta_s))
\nonumber \\
&\times&
s\left(t-\frac{R-R_s\cos(\theta-\theta_s)}c\right)
\label{received}
 \end{eqnarray}
 where $c$ is the speed of the light and $k=\omega_c/c$ where $\omega_c$ is the central radian frequency. We can therefore define the steering impulse response $h(z,t)$ for the $(R_s,\theta_s)$-user as
 \begin{eqnarray}
 h(z,t;s)
 &=&
 \exp(-jk(R-R_s\cos(\theta-\theta_s))
\nonumber \\
&\times&
\delta\left(t-\frac{R-R_s\cos(\theta-\theta_s)}c \right).
\label{impulse}
 \end{eqnarray}

The transmit pulse is chosen to be $s(t)=\sqrt{W}\text{sinc}(Wt)$ (where the $\text{sinc}(t)$ is here defined  as $\text{sinc}(t)=\sin(\pi t)/(\pi t))$
for the subsequent analysis, but any other waveform coming with a roll-off factor could be selected. 
Its correlation is therefore given by $\text{sinc}(Wt)$.
 Its bandwidth is denoted by $W$, and the associated space resolution, by $R_W=c/W$.

\subsubsection{Ambiguity function}

The ambiguity function is considered for the $(R_s,\theta_{s})$-UE and a second $(R_{\tilde{s}},\theta_{\tilde{s}})$-user. 
The  signal received at a $z$-AP from the $(R_s,\theta_{s})$-UE  is given by (\ref{received}).
We remind the reader that the $(R_s,\theta_{s})$-UE and the APs are supposed to be perfectly time and phase synchronous. 
The elements of the antenna array synchronously process the received signal by means of a combiner made of two operations: $t$-matched filtering and phase correction given by 
 \begin{eqnarray}
 r_s
 & =&   \int_{z=0}^{2\pi R} \exp(jk(R-R_s\cos(\theta-\theta_s))
 \nonumber \\
&\times& 
 \int_{-\infty}^{\infty} r(z,t;s) s^*\left(t- \frac{R-R_s\cos(\theta-\theta_s)
}c\right) dt \, R \, d\theta
 \nonumber \\
 &=&
2\pi R.
\label{steering6}
 \end{eqnarray}
If a $(R_{\tilde{s}},\theta_{\tilde{s}})$-UE  transmits, his/her contribution at the output of the $({R}_s, \theta_{s})$-matched combination would be given by equation (\ref{received}) with ${\tilde{s}}$ instead of $s$.
After $(R_s,\theta_{s})$-matched combiner, the resulting signal is
  \begin{eqnarray}
r(z;s,\tilde{s})
 &=&
 \exp(-jkR_{s\tilde{s}} \cos(z/R-\theta_{s\tilde{s}}) )
 \nonumber \\
&\times&\text{sinc}\left(\frac{R_{s\tilde{s}} \cos(z/R-\theta_{s\tilde{s}})}
{R_W}\right)
\label{steering7}
\end{eqnarray}
where we use $R_{s\tilde{s}} $ and $\theta_{s\tilde{s}}$ as the polar coordinates of the difference vector between vectors $(R_s,\theta_s)$ and $(R_{\tilde{s}},\theta_{\tilde{s}})$.
It appears that the $(R_{\tilde{s}},\theta_{\tilde{s}})$-UE contributes with a space-dispersive signal or a space-frequency selective spectrum, which has not been perfectly space-equalized by the $(R_s,\theta_{s})$-user matched filter.
The ambiguity function  is then given by
 \begin{eqnarray}
\text{AF}(R_{s\tilde{s}},\theta_{s\tilde{s}})
&=&\int_{z=0}^{2\pi R} 
 r(z;s,{\tilde{s}}) dz
 \nonumber \\
 &=&\int_{z=0}^{2\pi R} 
\exp(-jkR_{s\tilde{s}} \cos(z/R-\theta_{s\tilde{s}}))
\nonumber \\
&\times&
\text{sinc}\left(\frac{R_{s\tilde{s}} \cos(z/R-\theta_{s\tilde{s}})}{{R_W}}\right)
dz.
\label{ambig1}
\end{eqnarray}
It can be shown that, for an infinite number of APs, and a bandwidth $W=0$ or $R_W=\infty$,
 \begin{equation}
\text{AF}(R_{s\tilde{s}},\theta_{s\tilde{s}})
=2\pi R \, J_0(kR_{s\tilde{s}})
\label{gain0}
\end{equation}
where $J_0(.)$ is the  order-$0$-Bessel function of the first kind. The ambiguity function has zero crossings given by the roots of the Bessel function. The first zero is located at $kR_{s\tilde{s}}\simeq 2.4$, meaning that the resolution  can be defined by (there are several definitions of the resolution in the literature) by $R_{s\tilde{s}}=2.4/k=2.4\times \lambda/ 2 \pi \simeq 0.4 \times \lambda$. Hence the resolution is in the order of the wavelength ! However the Bessel function $J_0(.)$ also comes with sidelobes which are an issue, all the more when the number of antennas is finite. This will be discussed below.

\medskip 

About equation (\ref{ambig1}), we have that both functions inside the integral are periodic functions of $z$ with period $2\pi R$.
We can therefore use their respective Fourier series expansion, which actually provides their spectral description.
For the $\exp(.)$ term, we use the Jacobi-Anger expansion which leads to 
\begin{eqnarray}
\exp[-jk(R_{s\tilde{s}} \cos(z/R-\theta_{s\tilde{s}}))]
&=&
\nonumber \\
\exp[-jkR_{s\tilde{s}} \sin(z/R+\pi/2-\theta_{s\tilde{s}}))]
&=&
\nonumber \\
\sum_{n=-\infty}^{\infty}
\exp(jn(\pi/2-\theta_{s\tilde{s}})) \, J_n(-kR_{s\tilde{s}} ) \exp(jnz/R).
\end{eqnarray}
We can also use a Fourier series expansion for the $\text{sinc}(.)$ function. By analogy with the Bessel functions, we use the following expansion~:
\begin{equation}
\text{sinc}\left(\frac{R_{s\tilde{s}} \sin(z/R)}{{R_W}}\right)
=
\sum_{l=-\infty}^{\infty}
L_l\left(\frac{R_{s\tilde{s}}}{{R_W}}\right) \exp(jlz/R )
\nonumber
\end{equation}
with
\begin{equation}
L_l\left(\frac{R_{s\tilde{s}}}{R_W}\right)
=
\frac 1{2\pi } \, \int_{\theta=0}^{2\pi } \, \text{sinc}\left(\frac{R_{s\tilde{s}} \sin(\theta)}{{R_W}}\right)
 \exp(-jl\theta)d\theta.\nonumber
 \end{equation}
 It then comes
 that 
 \begin{eqnarray}
\text{sinc}\left(R_{s\tilde{s}} \sin\left(\frac{z/R-\cos(z/R-\theta_{s\tilde{s}})}{{R_W}}\right)\right)
&=&
\nonumber \\
\sum_{l=-\infty}^{\infty}
\exp(jl(\pi/2-\theta_{s\tilde{s}}))
L_l\left(\frac{R_{s\tilde{s}}}{{R_W}}\right) \exp(jlz/R).
\nonumber
\end{eqnarray}

Plugging everything together, the ambiguity function can be shown to be (the derivations are left out for the sake of concision)
 \begin{eqnarray}
\text{AF}(R_{s\tilde{s}},\theta_{s\tilde{s}})
&=&2\pi R \sum_{n=-\infty}^{\infty}
\, J_n(-kR_{s\tilde{s}} )
 L_{n}\left(\frac{R_{s\tilde{s}}}{R_W}\right).
 \label{gainax}
\end{eqnarray}
While equation (\ref{ambig1}) shows the computation of the ambiguity function by means of a correlation in the $z$-space domain, equation (\ref{gainax}) actually shows the equivalence with a correlation in the spatial $k_z$ frequency domain (Fourier series coefficients), a result known as the Parseval identity. The  harmonics located in $k_{z}=n/R$ or $k_{\theta}=k_{z}R=n$ ($n$ integer)  are actually the spectral components of the ambiguity function. If the signal is really truncated for $z\in[0,2\pi R]$ the frequency continuous spectrum can be obtained by interpolating the Fourier series coefficients by means of a $\text{sinc}(k_zR/2)$ waveform. 
As a sanity check, it appears that this result boils down to equation (\ref{gain0}) for ${R_W} \rightarrow \infty$.
It is also interesting to notice that the $ L_{n}(R_{s\tilde{s}}/{R_W})$ terms will help attenuate the sidelobes of the $J_n(-kR_{s\tilde{s}} )$ terms. This shows the advantage of using a non-zero bandwidth, with an ${R_W}$ value as small as possible, meaning a large bandwidth $W$.
\begin{figure}[hbt]
\centerline{\includegraphics[width=3.5in]{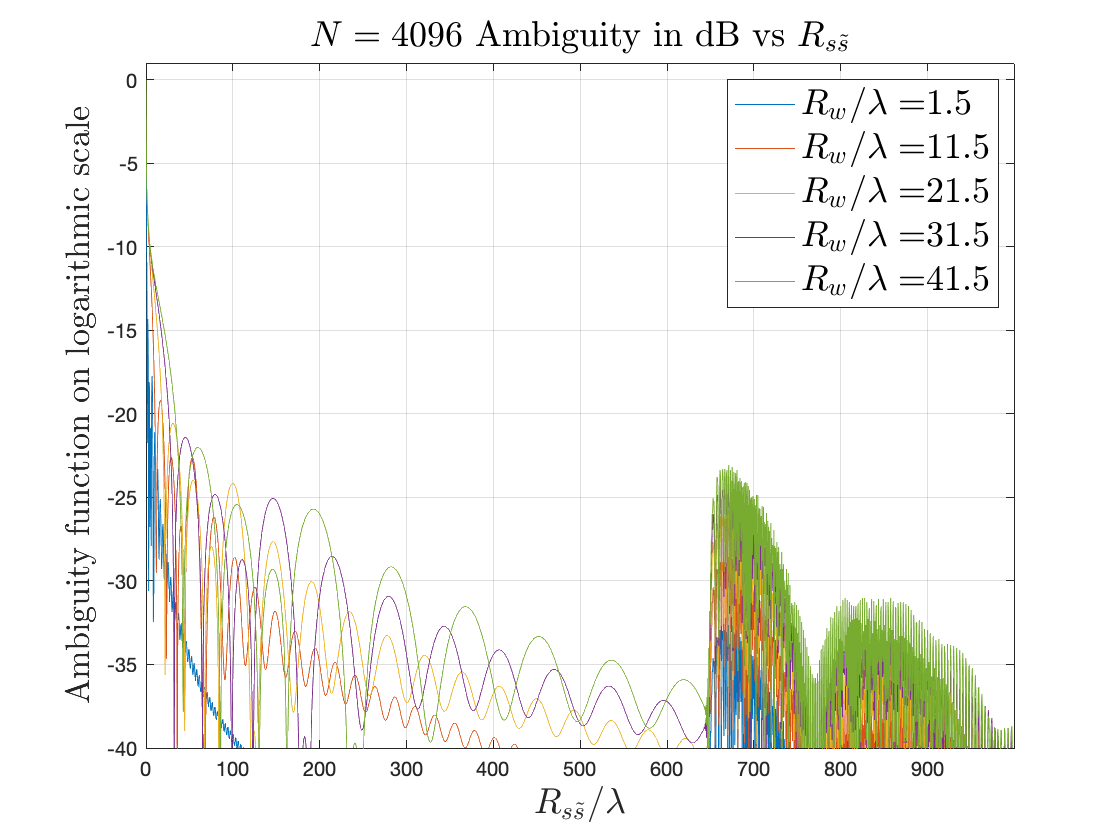}}
\caption{Ambiguity function for a cell-free configuration in dB, versus $0 \leq
R_{s\tilde{s}}/\lambda \leq 1000$ for various values of ${R_W}$ which also means $W$: ${{R_W}}/\lambda= 1.5, 11.5, 21.5, ...$. The number of antennas is $N=4096$.
.\label{AF_4096_1000_070824}}
\end{figure}

\subsubsection{Finite number of antennas}

We now consider the case where the number of regularly spaced antennas is limited to $N$, which means sampling along $z$, the radial variable. With sampling in the $z$ domain, aliasing will occur in the corresponding frequency domain, that is the domain of $k_z$. With $z=n \Delta z$ where $n\in{\cal N}^+$, as mentioned above, the sampling period $\Delta z$ gives rise to repetitions of the spectrum along multiples of the frequency $k_z=2\pi/\Delta z$. If we want to cover a maximum frequency $k_{z,\text{max}}$, we must have $2\pi/\Delta z>2k_{z,\text{max}}$. In the cell-free case, $k_{z,\text{max}}=kR_{s,\text{max}}/R$ as explained above. Hence, we must have $\Delta z<\pi/k_{z,\text{max}}=\pi R/kR_{s,\text{max}}$, leading to $\Delta z<\lambda R/2R_{s,\text{max}}$. 
As $\Delta z=2\pi R/N$, it comes $\Delta z=2\pi R/N<\lambda R/2R_{s,\text{max}}$ or $N>4\pi R_{s,\text{max}}/\lambda $. For a given $N$, the spectrum gets repeated along multiples of  $k_z=2\pi/\Delta z=N/R$. The corresponding $R_{\text{s,max}}=N\lambda /4\pi$. Any $R_s>R_{\text{s,max}}=N\lambda /4\pi$ will  be responsible for aliasing and corruption of the spectrum of interest, or so to say, a contribution corresponding to unwanted $(R_{\tilde{s}},\theta_{\tilde{s}})$-users.
We can simply use the results obtained above about the Fourier series expansions. 
Plugging everything together, we get an ambiguity function
 \begin{eqnarray}
\text{AF}(R_{s\tilde{s}},\theta_{s\tilde{s}})
&=&\sum_{n=-\infty}^{\infty}
\sum_{l=-\infty}^{\infty}
\exp(jn(\pi/2-\theta_{s\tilde{s}})) \, 
\nonumber \\
&\times&
\exp(jl(\pi/2-\theta_{s\tilde{s}}))
J_n(-kR_{s\tilde{s}} )
L_l\left(\frac{R_{s\tilde{s}}}{{R_W}}\right) 
\nonumber \\
&\times&
\sum_{i=0}^{N-1} 
 \exp(2\pi j i n/N) \exp(2\pi j i l/N)
 \nonumber \\
&=&
N\sum_{n=-\infty}^{\infty} \sum_{p=-\infty}^{\infty}
\exp(jpN(\pi/2-\theta_{s\tilde{s}})) 
\nonumber \\
&\times&
\, J_{n+pN}(-kR_{s\tilde{s}} )
 L_{n}\left(\frac{R_{s\tilde{s}}}{R_W}\right).
 \label{gainaxdiscret2}
\end{eqnarray}
The summation over $p$ is responsible for aliasing terms.
These results are illustrated by Fig.\ref{AF_4096_1000_070824}. The figure provides the ambiguity function (\ref{gainaxdiscret2}) in dB as a function of 
$0 \leq
R_{s\tilde{s}}/\lambda \leq 1000$ for various values of ${R_W}$ and a number of antennas $N=4096$. The values of ${R_W}/\lambda=1.5, 11.5, 21.5, ..., ...$, meaning a decreasing bandwidth. The value of the phase error is again arbitrarily set to $\theta_{s\tilde{s}}=3\pi/37$. 
This time, the ambiguity function remains a function of $\theta_{s\tilde{s}}$. The impact of $\theta_{s\tilde{s}}$ will be all the more important when $N$ decreases.
The aliasing along the radial axis $(R_{s\tilde{s}}/\lambda)$ associated with the finite number of antennas is clearly visible. For a selected $R_{\text{s,max}}=N\lambda /4\pi$, the ambiguity function will repeat itself around $2R_{\text{s,max}}=N\lambda /2\pi$. For $N=4096$, $2R_{\text{s,max}}=N\lambda /2\pi=651$. The aliasing around the correct $2R_{\text{s,max}}$ is visible on the figure.  
As the blue curve confirms, it again clearly appears that the presence of a large $W$ (small ${R_W}$) waveform helps drastically reduce the sidelobes of the ambiguity function and the effect of aliasing. In this plot, the summations in equation (\ref{gainaxdiscret2}) are truncated: $l \leq 20$ and $p \leq 5$. 
\begin{figure}[hbt]
\centerline{\includegraphics[width=3.5in]{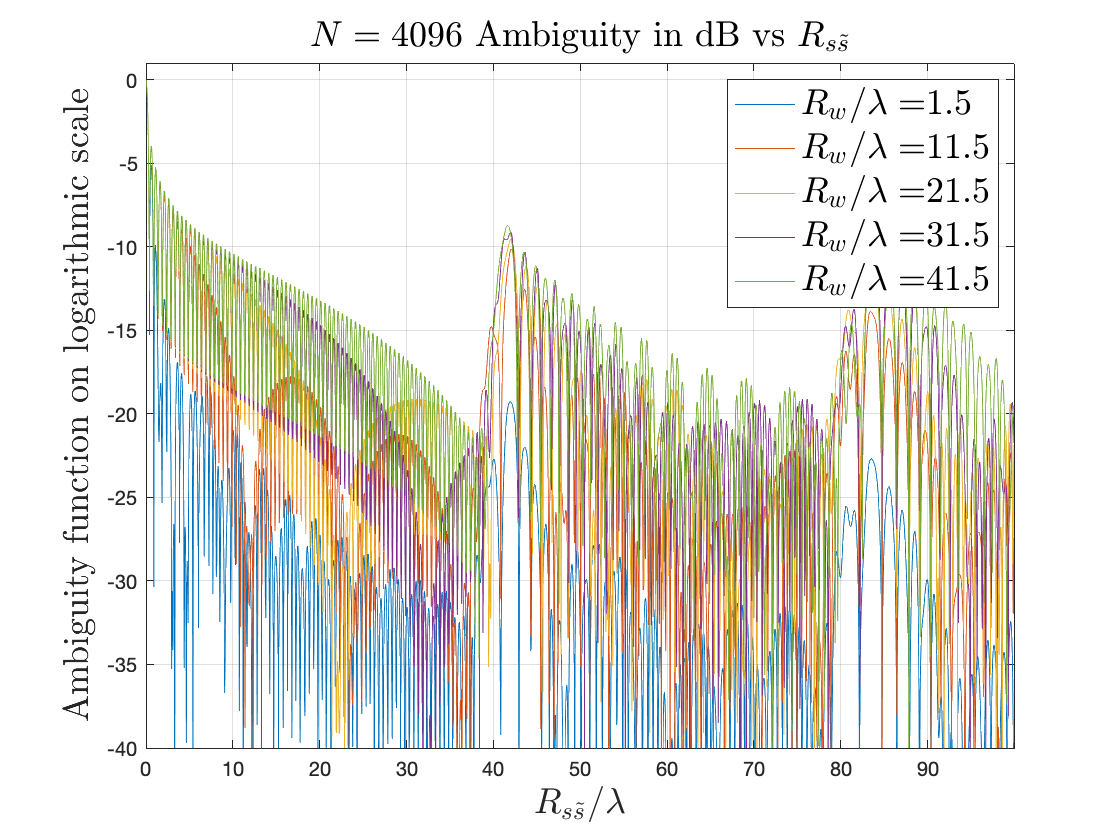}}
\caption{Ambiguity function for a cell-free configuration in dB, versus $0 \leq
R_{s\tilde{s}}/\lambda \leq 100$. ${R_W}/\lambda=1.5, 11.5, 21.5, ...$. The number of antennas is $N=256$. $\theta_{s\tilde{s}}=3\pi/37$.
\label{AF_256_100_070824}}
\end{figure}
Fig. \ref{AF_256_100_070824}
 provides the ambiguity function (\ref{gainaxdiscret2}) this time for a  number of antennas $N=256$ and $0 \leq
R_{s\tilde{s}}/\lambda \leq 100$. All other parameters are the same as above. The reduced number of antennas means that the largest radius inside which users can be positioned is reduced compared to the radius permitted by
$N=4096$. 

\subsection{Communications}

\subsubsection{Transmit beampointing}

When transmit beampointing is implemented, the measure of its efficiency is given by the array gain. This gain is obtained when all APs synchronously transmit to a target UE. If the $(R_s,\theta_{s})$-user  is again the target UE, interference can be measured at a $(R_{\tilde{s}},\theta_{\tilde{s}})$-user. The link between the AP located in $z$ and the $(R_s,\theta_{s})$-user still has an impulse response given  by (\ref{impulse}). This impulse response indicates that the signal generated by an AP towards an $(R_s,\theta_{s})$-user $(R_{\tilde{s}},\theta_{\tilde{s}})$-user will have time delays which depend on $z$. This means that the channel is time dispersive, with a dispersion depending on $z$.
 An MRT precoder compensates the phase and the delay, and transmits
 \begin{equation}
 r_t(z,t;s)
 =
 \exp(jkR_s\cos(\theta-\theta_s))
s\left(t+\frac{R_s\cos(\theta-\theta_s)}c\right).
\nonumber
 \end{equation}
 UE $(R_s,\theta_{s})$ receives, after transmission over the channel a signal whose value or its array associated array gain is simply equal to $2\pi R$.
 For a UE  located in $(R_{\tilde{s}},\theta_{\tilde{s}})$, after propagation over the channels, its signal will add up with the $(R_s,\theta_{s})$-UE's one. After user-$(R_s,\theta_{s})$ delay and phase correction the AP side, the array gain for the $(R_{\tilde{s}},\theta_{\tilde{s}})$-user will be given by
 
\begin{eqnarray}
G(R_{s\tilde{s}},\theta_{s\tilde{s}})
&=&  \int_{z=0}^{2\pi R} \,
\exp(-jkR_{s\tilde{s}} \cos(z/R-\theta_{s\tilde{s}}))
\nonumber \\
&\times& \text{sinc}\left(\frac{R_{s\tilde{s}} \cos(z/R-\theta_{s\tilde{s}})}{R_W} \right)
 \end{eqnarray}
which corresponds exactly to equation (\ref{ambig1}). It means that when the signal originating from the AP is shaped for the $(R_s,\theta_{s})$-user, the $(R_{\tilde{s}},\theta_{\tilde{s}})$-user still sees a space dispersive signal which has not been properly pre-equalized.
Hence the array gain shows its similarities with the ambiguity function and delivers thereby information about the capability to serve neighboring users with limited, or not, interference. The role played in positioning by the user of interest in  $(R_s,\theta_{s})$ is here the UE for which the MRT is designed. The $(R_{\tilde{s}},\theta_{\tilde{s}})$-UE which interferes with the UE of interest in positioning is here played by the UE interfered in the DL. 
The impact of the bandwidth $W$ on ambiguities reductions, and on the aliasing when a finite number of antennas is used, also applies for a DL. 
The number of antennas $N$ will impact the distance over which aliasing is generated, and, thereby interference over users. 
The interference generated by the aliasing is larger than that associated with the sidelobes of the $J_0$ contribution closer to  $(R_{\tilde{s}}=0)$.

\subsubsection{Receive combining}

In an UL, the operation performed by the MRC designed for UE $(R_s,\theta_{s})$ produces a signal which is exactly the ambiguity function derived above when a $(R_{\tilde{s}},\theta_{\tilde{s}})$-UE is transmitting at the same time. The number of antennas impacts the number of additional users seen because of the aliasing and that will interfere with the targeted UE.

\section{Conclusion}

This paper has considered a UCCFN with antennas distributed on a circle around target users. For positioning, the ambiguity function has been derived as a function of the distance between a reference and an interfering user. When a finite number of antennas is used, space aliasing appears which translates into the fact that unwanted users may create an interference on the reference user signal. It has been shown how the bandwidth of the transmitted signal helps reduce the
aliased versions. For instance, for $N=256$, the aliasing is attenuated by $10$dB or more. It has also been shown that a completely parallel analysis can be conducted for data transmission with MRT or MRC. For these cases, we obtained the array gains and discussed the impact of the bandwidth and of the number of antennas on interference between users. The theoretical results have been illustrated and corroborated by computational results.

\medskip

These results  also impact, on the one hand, the resolution with which scatterers can be observed, and, on another hand, the granularity with which the space should be sampled to generate relevant steering vectors. This will be investigated in future work.

\end{document}